\documentstyle[preprint,aps]{revtex}
\begin{document}
\draft
\title{Random bond Ising chain in a transverse magnetic field:\\
       A finite-size scaling analysis.}
\author{A. Crisanti}
\address{Dipartimento di Fisica, Universita di Roma ``La Sapienza'',
         00185 Roma, Italy}
\author{H. Rieger}
\address{Institut f\"ur Theoretische Physik, Universit\"at zu K\"oln,
         50937 K\"oln, Germany}
%
\date{May 27, 1994}
\maketitle
\begin{abstract}
We investigate the zero-temperature quantum phase transition of the 
random bond Ising chain in a transverse magnetic field. Its critical 
properties are identical to those of the McCoy-Wu model, 
which is a classical Ising model in two dimensions with layered disorder.
The latter is studied via Monte Carlo simulations and transfer matrix
calculations and the critical exponents are determined with a finite-size 
scaling analysis. The magnetization and susceptibility obey conventional
rather than activated scaling. We observe that the order parameter--
and correlation function--probability distribution show a nontrivial
scaling near the critical point which implies a hierarchy of critical
exponents associated with the critical behavior of the generalized 
correlation lengths.
\end{abstract}
Quite recently there has been a growing interest in the
zero-temperature quantum critical behavior of {\it disordered} spin
systems. Thermal fluctuations are absent here and the phase transition 
is driven by the interplay between randomness and quantum fluctuations.  
In order to tune the system to criticality one can either vary the 
strength of the disorder, as for instance in spin-$\frac12$ XXZ--chains 
\cite{XXZ}, or one can control the strength of the quantum fluctuations
directly by an external transverse magnetic 
field in spin models with a strong Ising anisotropy. The latter case is 
particularly interesting since experimentalists became able to investigated 
the effect of a transverse field on the glass transition in the Ising spin 
glass LiHo$_x$Y$_{1-x}$F$_4$ at low temperatures\cite{exp}. On the
theoretical side much progress has been made since then:
the infinite range model has been solved analytically 
\cite{mft2,mft1,oppermann},
a Migdal-Kadanoff renormalization group calculation has been made
\cite{Santos}, the critical
exponents in 2 and 3 dimensions have been determined via Monte Carlo 
simulations \cite{qsg2d,qsg3d} and new results for the one-dimensional
case have been derived via a renormalization-group analysis \cite{DF}. 

In the latter papers focusing on the critical behavior in finite 
dimensions of these quantum models it has been pointed out that 
their universal properties are identical to those of classical 
Ising models with layered disorder \cite{lev}. Especially the Ising spin
chain in a transverse field can be mapped onto the McCoy-Wu
model \cite{mccoy,zittartz,shankar}, for which various exact 
results have been derived. The critical exponents of 
the order-parameter $\beta$ and the correlation length $\nu$ as
well as the dynamical exponent $z$ has been found recently via a 
renormalization-group (RNG) analysis by D.\ Fisher \cite{DF}.

The aim of the present paper is to perform a numerical investigation
of the finite-size scaling behavior of the Ising spin chain in a
transverse magnetic field. Such an analysis has not been performed
yet for this model and bears some new features concerning the finite-size 
scaling of anisotropic systems \cite{binder}. It can also be
seen as a test-ground for numerical methods applied to systems, for
which --- in contrast to this model --- no quantitative theoretical 
prediction are at hand (e.g.\ for the cases considered in
\cite{qsg2d,qsg3d}). And finally it provides a check to what extend
analytical predictions, like those made in \cite{DF} and which are valid
asymptotically for an infinite system with rather unusual properties,
can be detected in systems of finite size.

The model under consideration is described by the quantum Hamiltonian
\begin{equation}
H_Q=-\sum_i J_i\sigma_i^z\sigma_{i+1}^z-
\Gamma\sum_i\sigma_i^x\;,\label{eq1}
\end{equation}
where $\sigma$ are spin--$\frac12$ Pauli matrices,
$\Gamma$ is the transverse field strength and the exchanges $J_i$
are quenched random variables obeying a distribution $P(J)$.
At zero temperature the system (\ref{eq1}) has a ferromagnetic
phase transition to long--range magnetic order at a critical value 
$\Gamma_c$, which depends on the bond distribution $P(J)$. 
We are interested in the critical properties
of this transition. However, as shown in \cite{DF,mccoy,shankar},
the magnetization behaves already non-analytically at higher values of
$\Gamma_c$ giving rise to e.g.\ a divergence of the longitudinal 
susceptibility at higher values of $\Gamma$.

The ground state energy of this one-dimensional quantum model 
(\ref{eq1}) is equal to the free energy of the two-dimensional classical
Ising model \cite{suzuki}
\begin{equation}
H=-\sum_{i,j}\tilde{J}_i S_{i,j}S_{i+1,j}-K\sum_{i,j}S_{i,j}S_{i,j+1}
\label{eq2}
\end{equation}
at a certain finite temperature $T$. 
Here $S_{i,j}=\pm1$ are classical Ising spins, the site index $i$ 
runs along the $x$ (space) direction and the index $j$ along the 
$\tau$ (imaginary time) direction of a two-dimensional square lattice.
Following \cite{qsg2d,qsg3d} we can rescale the bond strengths
$\tilde{J_i}$ and coupling constant $K$ without changing the universal
properties. For numerical convenience we set $K=1$ and take a binary 
distribution 
\begin{equation}
P(\tilde{J})=\frac12 \delta(\tilde{J}-j_1)+\frac12
\delta(\tilde{J}-j_2)\;,\label{eq3}
\end{equation}
where we put $j_1=1$.
The layered random bond Ising model (\ref{eq2}) with the bond
distribution (\ref{eq3}) has a ferromagnetic 
phase-transition at a critical temperature $T_c$ defined by \cite{mccoy}
\begin{equation}
\log\,\coth\,(1/T_c)+\log\,\coth\,(j_2/T_c) = 4 / T_c
\label{eq:Tc}
\end{equation}
and the universal properties (like exponents etc.) are identical to
those of the quantum chain (\ref{eq1}) at $\Gamma_c$ and zero
temperature. Therefore we study model (\ref{eq2}) at $T_c(j_2)$ by
Monte-Carlo simulations of
rectangular lattices of size $L\times L_\tau$. The largest size
in the $\tau$-direction was $L_\tau=160$ whereas that in 
the space direction $x$, which corresponds to the length of 
the quantum chain (\ref{eq1}), was $L=16$. Hence the
disorder average over the distribution (\ref{eq3}) could be done 
exactly by generating all, non-equivalent bond-configurations
(whose number is approximately $2^L/L$).
To have more confidence on the data from the Monte-Carlo simulation we
have compared our results to those obtained from transfer matrix
calculations. The advantage of the latter method is that the results are
exact, but the drawback is that one is limited to small system sizes $L\le10$.
In all cases we found no significant deviations between the results of the 
two methods.

The correlation length of the quantum chain (\ref{eq1})
diverges as $\xi\sim(\Gamma-\Gamma_c)^{-\nu}$ when approaching
the ferromagnetic transition.
The characteristic relaxation time of the quantum dynamics
is expected to diverge as $\tau\sim\xi^z$, with $z$ being the dynamical
exponent. These two diverging scales can naturally be found in the
classical model (\ref{eq2}): due to the extreme anisotropy one expects 
the correlation length in the space (or $x$) direction to 
diverge like $\xi\sim(T-T_c)^{-\nu}$ and the correlation length in the
imaginary time (or $\tau$) direction like $\xi_\tau\sim(T-T_c)^{-z\nu}$.
Following a nice argument by L.\ Mikheev \cite{private}
one might imagine the system close to $T_c$ as being 
composed of roughly rectangular, ferromagnetic/paramagnetic domains, 
which are located (in the space direction) at segments of the chain 
where strong/weak bonds are dominating. The correlation length $\xi_\tau$
in the time direction is then given by the average distance of domain walls in
semi-infinite strips of width $\xi$ being ferromagnetically ordered, 
thus $\xi_\tau\propto\exp(a\xi)$ (see e.g.\ \cite{kleban}). This
supports very much an activated dynamics scenario with $z=\infty$,
as found in the RNG analysis \cite{DF}.
However, we begin the analysis by assuming a finite $z$ and will
see how far we get.

{\it At} the critical point (i.e.\ $T=T_c$) various thermodynamic
quantities are expected \cite{binder} to depend only on the scaling 
variable $L/L_\tau$, the aspect ratio or the shape of the system.
For instance we would have for the averaged spontaneous magnetization
\begin{equation}
M=[\langle m\rangle]_{\rm av}
\approx L^{-\beta/\nu}\widetilde{m}(L_\tau/L^z)\;,
\label{mag}
\end{equation}
where $m=(L_\tau L)^{-1}\bigl|\sum_{i,j}S_{i,j}\bigr|$ is
the magnetization per site,
$\langle\cdots\rangle$ means the thermal average and
$[\cdots]_{\rm av}$ means the disorder average. In figure 1 we show
a scaling plot according to (\ref{mag}) obtained by Monte-Carlo
simulations for different shapes and sizes (note that the disorder
average is done exactly) at $j_2=0.1$ and $T=T_c=1.32038$. It yields
$\beta/\nu=0.17\pm0.01$ and $z=1.65\pm0.05$. We also looked at
$j_2=0.05$ for which $T_c=1.14710$, and obtain $z=1.70\pm0.05$ and 
$\beta/\nu=0.18\pm0.01$.
The RNG prediction is $\nu=2$ (for the averaged correlation length)
$\beta=(3-\sqrt{5})\approx0.38$ \cite{DF}, which yields a ratio
$\beta/\nu\approx0.19$ that agrees roughly with our estimate.

The exponent $z$ decreases as $j_2$ increases 
approaching $z=1$ for $j_2=1$ and indicating the 
crossover to the pure case. This is the reason why we used small
values for $j_2$ to ensure that what we see is the critical behavior
of the disordered model. On the other hand we tried to avoid too small
values of $j_2$ in the Monte-Carlo simulations, since then the critical
temperature decreases too much and equilibration becomes more difficult.

The insert of figure 1 shows the scaling behavior of the susceptibility
at $j_2=0.1$,
\begin{equation}
\chi(L,L_\tau) = L_\tau L [\langle m^2\rangle]_{\rm av}
\approx L^{\gamma'/\nu}\tilde{\chi}(L_\tau/L^z)\;,
\label{sus}
\end{equation}
for which we find $\gamma'/\nu=2.3\pm0.1$ (we used the value $z=1.65$
that was obtained from the scaling of the magnetization).
Note that $\gamma'$ is
not the critical exponent that describes the divergence of the 
susceptibility in an infinite system by approaching the temperature 
$T_c$ from above, since this quantity is expected to diverge already at a 
temperature higher than $T_c$ \cite{DF,mccoy}. The prediction
for an anisotropic system obeying hyperscaling \cite{binder} is
$\gamma'/\nu+2\beta/\nu=d+z$. Inserting $d=1$ and the values for 
$\beta/\nu$ and $z$ given above this relation is fulfilled very well.
For the magnetization and the susceptibility
conventional scaling seems to work well for these system sizes.

The averaged cumulant 
$g_{\rm av}=0.5\cdot[3-\langle m^4\rangle/\langle m^2\rangle^2]_{\rm av}$
is expected to scale like 
\begin{equation}
g_{\rm av}(L,L_\tau)=\widetilde{g}(L_\tau/L^z)
\label{cum}
\end{equation}
and a scaling 
plot is shown in figure 2 with $z=1.55\pm0.05$ for $j_2=0.1$, which
is slightly lower than the estimate from the spontaneous magnetization.
Furthermore the data collapse is not as good as in figure 1. Even worse
is the scaling behavior of the cumulant 
$\overline{g}=0.5\cdot(3-[\langle m\rangle^4]_{\rm av}/
[\langle m\rangle^2]_{\rm av}^2)$.
A systematic shift in the maximum to smaller values for increasing system
sizes $L$ indicates that this quantity is not dimensionless as expected
naively. The natural scaling assumption 
$[\langle m\rangle^k]_{\rm av}\approx L^{-k\beta/\nu}\widetilde{m}_k(L_\tau/L^z)$
does not seem to be correct here. As a consequence,
the order parameter probability distribution 
$P({\cal M})=[\delta(\langle m\rangle-{\cal M})]_{\rm av}$
does not scale in a trivial way like $P({\cal M})\approx 
L^{\beta/\nu}\widetilde{P}({\cal M} L^{\beta/\nu},L_\tau/L^z)$
--- as it does e.g.\ in conventional spin glasses \cite{bhayou} ---
which we checked explicitly by looking at the magnetization
histograms for systems with constant aspect ratios $L_\tau\approx L^z$.

To have an independent check of this scenario we have compared the above 
results with those obtained from the transfer matrix calculation.
We have used the method recently introduced in \cite{CPSV} for the exact 
calculation of the free energy derivatives. The method has been extended to
a finite rectangular lattice with periodic boundary condition in both 
directions. 
By expressing the cumulants of the magnetization as derivatives 
of the free energy, we have computed $g_{\rm av}$ and 
$\overline{g}$ for systems sizes up to $8\times 256$. 
Since there are no numerical derivatives involved and the average over 
disorder is performed by summing over all possible configurations, 
the calculation yields the exact values. In all cases we did not find 
significant difference between these results and those of the Monte-Carlo 
simulations.

We have also calculated the averaged spin-correlation function at $T_c$, 
which is defined as
\begin{equation}
C(r,t)=[\langle S_{i,j} S_{i+r,j+t}\rangle]_{\rm av}\;.
\end{equation}
For the averaged correlations in the time direction $C(0,t)$
one expects \cite{binder} for $L_\tau\propto L^z$ a behavior
\begin{equation}
C(0,t)\propto t^{-\eta_\perp}+(L_\tau-t)t^{-\eta_\perp}\;,
\label{corrtime}
\end{equation}
where the second term on the r.h.s.\ takes into account the periodic
boundary conditions. In the insert of figure 3 we have depicted $C(0,t)$
for various system sizes $L$ with $L_\tau$ chosen at the maximum of
$g_{\rm av}(L_\tau)$, so that $L_\tau\propto L^z$. From the fit
we conclude that $\eta_\perp=0.23\pm0.01$. 

Concerning the spatial correlation function $C(r,0)$
Shankar and Murthy report a result
(see equations 3.39 and 3.43 in \cite{shankar}), which is
\begin{equation}
C(r,0)\propto\exp(-a\sqrt{r})+\exp(-a\sqrt{L-r})\;,
\label{stretch}
\end{equation}
where the second term on the r.h.s.\ takes again into account the
periodic boundary conditions. This form 
yields a nice least square fit to our numerical data, as shown in 
figure 3. Note that comparing (\ref{corrtime}) and (\ref{stretch})
one observes that ``space'' and ``time'' seem to scale
like $r\sim(\log t)^2$, as predicted in \cite{DF}.

D.\ Fisher argues \cite{DF} that the result (\ref{stretch}) holds for
the {\it typical} correlations (i.e.\ those corresponding to
the maximum of their probability distribution), whereas the average 
should decay algebraically, since the latter should be dominated by
rare, strongly correlated regions of the spin chain.
We tried to fit $C(r,0)$ to an algebraic decay similar to
(\ref{corrtime}) with a different exponent $\eta_\parallel$, which
gave much worse results. However, stipulating %
\begin{equation}
C(r,0)\approx r^{-\eta_\parallel}\widetilde{c}(r/L\,,\,L_\tau/L^z)\;,
\label{alg}
\end{equation}
the relation 
$C(L/2,0)\approx L^{-\eta_\parallel}\overline{c}(L_\tau/L^z)$
should hold, which gives indeed an acceptable data-collapse for
$\eta_\parallel=0.40\pm0.02$ (using our estimate $z=1.70$). 
The latter result agrees well 
with the result $\eta_\parallel\approx0.38$ obtained in \cite{DF}.
Furthermore it is
consistent with the scaling relation $\eta_\perp/\eta_\parallel=z$
(within the errorbars) and $\eta_\parallel=d+z-\gamma'/\nu=2\beta/\nu$.
We would like to stress that our data are compatible with both equations
(\ref{stretch}) and (\ref{alg}), which, however, are based on assumptions
excluding each other.

Next we study the probability distribution of correlation functions in 
the spatial direction through the analysis of the associated generalized
correlation lengths \cite{Derrida,CNPV}. 
The analysis is carried out using the transfer matrix 
approach, for more details see \cite{CNPV,crisanti}.
We focus on the spatial correlation function 
for which we can use the general results for products of random 
matrices. In this case the latter consists of a succession
of transfer-matrices from row $i$ to row $i+1$ in the spatial direction,
each of which is made by $L_{\tau}$ spins. Since the generalized 
correlation lengths to be defined below are related to the
first two Lyapunov exponents of an infinite product of transfer
matrices we have to choose $L$ very large ($\sim10^6$). An
advantage of this approach is that one is left with only one
scaling variable at the critical point since $L_\tau/L^z$ is zero.

In general each moment of the probability distribution of the 
correlation function defines
a characteristic length scale which we denote by $\xi_q$ where $q$ is the
order of the moment \cite{Derrida,CNPV}:
\begin{equation}
  \xi_q^{-1} = -\lim_{r\to\infty} \frac{1}{rq}\, \ln
                     \left[ G_i(r)^q \right]_{\rm av}
\label{eq:genCL}
\end{equation}
where $G_i(r)$ is the connected correlation function between the row $i$
and the 
row $i+r$. For example $\xi_1$ is the characteristic length scale
of the average correlation function, while
$\xi_0$ is that of the typical correlation function. We shall then
call $\xi_0$ the typical correlation length and $\xi_1$ the average
correlation length. It can be shown that if $q>q'$ then 
$\xi_q \geq \xi_{q'}$ \cite{crisanti}. 
Taking into account this hierarchy, the usual finite-size 
scaling hypothesis \cite{binder} for the spatial generalized correlation 
lengths would be
\begin{equation}
 \xi_q^{-1} = L_{\tau}^{-1/z_q}\, 
                  \widetilde{\zeta}_q ( (T-T_c)\, L_{\tau}^{1/z_q\nu_q} )
\label{eq:scl}
\end{equation}
where the $z_q$ can be called ``generalized dynamical exponents''. From 
the relation $\xi_q \geq \xi_{q'}$ for $q>q'$ it follows that 
$z_q \geq z_{q'}$ and $\nu_q \geq \nu_{q'}$. 
In the scaling form (\ref{eq:scl}) 
it is assumed that the $\xi_q$ diverges all at the same temperature $T_c$,
which in general need not to be the case, see for instance \cite{Derrida}.
As consequence, especially for large $q$ (\ref{eq:scl}) should be 
modified allowing for a $q$-dependent $T_c$. However, we stick to 
the scaling form (\ref{eq:scl}) with $T_c$ given by the
ferromagnetic phase transition temperature.

We have calculated (for technical details see \cite{CNPV,crisanti})
the $z_q$ by computing the exponents $\xi_q$ at the
critical point $T_c$ for different $j_2$ for systems of sizes up
to $L_{\tau}=8$. The length of the product was $10^6$. The exponent $z_q$ 
increases as $j_2$ decreases. We get (with at least 10 percent
accuracy)
\begin{equation}
\begin{array}{l|ccc}
 j_2 & z_0 & z_1 & z_2 \\
\hline
0.10 & 1.74 & 2.04 & 2.63 \\
0.05 & 2.12 & 3.23 & 5.26 \\
\end{array}
\label{exponents}
\end{equation}
In all cases we found a good scaling at the critical temperature $T_c$ 
given by (\ref{eq:Tc}). Our statistics is not accurate enough to 
investigate higher correlation lengths. The dynamical exponents
increase systematically with decreasing $j_2$ and it cannot be ruled out 
that $z_q\rightarrow\infty$ for $j_2\rightarrow0$. Since the other exponents
(as $\nu_q$, see below, or $\beta$ see above) are less
susceptible to a variation in $j_2$ we have to leave it open here,
whether this behavior indicates a cross-over or an actual non-universality
of $z_q$ with respect to $j_2$.

The exponent $\nu_q$ are obtained from the data collapse using 
(\ref{eq:scl}). In this case we computed the generalized correlation 
lengths for temperatures $T>T_c$ and for system sizes up to $L_{\tau}=7$. 
Again the length of the product was $10^6$. With this statistics we 
were able to estimate only the first two exponents. The data for
$j_2=0.1$, $0.05$ and $0.01$ leads to $\nu_0\simeq 0.7$ and 
$\nu_1\simeq 1$, in agreement with the analytical result 
of \cite{shankar} but disagrees with the RNG result $\nu_1=2$ \cite{DF}.
In figure 4 we show a scaling plot for $\xi_0$.

We briefly discuss the time-correlation length for which we consider
a semi-infinte strip in the time-direction ($L_\tau\rightarrow\infty$).
While for the spatial correlations we have an infinite product of
random matrices, in the case of time-correlations the transfer matrix
is always the same since the randomness is only in the spatial direction. 
In this situation for each realization of disorder the correlation length
is given by the inverse of difference between the first two eigenvalues 
of the transfer matrix and yields, if averaged, the inverse of the typical 
correlation length, $\xi_0^{-1}$ (note that  this is equivalent to averaging 
the logarithm of the correlation function).
For the typical time-correlation length we stipulate again
the usual finite-size scaling form \cite{binder}
\begin{equation}
 \xi_{\tau,0}^{-1} = L^{-\tilde{z}_0}\, 
                  \widetilde{\zeta}_{\tau,0} ( (T-T_c)\, L^{1/\tilde{\nu}_0} ).
\label{eq:scltime}
\end{equation}
For $j_2=0.1$ we found, for example, $\widetilde{z}_0\approx1.3$ and 
$\widetilde{\nu}_0\approx1$. Note that $\tilde{z}_0\ne z_0$ and
$\widetilde{\nu}\ne\nu$.

To summarize we performed a detailed finite-size scaling analysis
of the zero-temperature phase transition occurring in a random
bond Ising chain by tuning the transverse magnetic field to some
critical value. For this model many analytical results are known
and our analysis shows a good agreement with the results
of Shankar and Murthy \cite{shankar} and concurs also with the 
RNG prediction of the existence of 
different length scales with different critical exponents $\nu$ \cite{DF}.
However, we do not find the same values as those reported in the latter.
One possible explanation of this fact might be the following:
D.\ Fisher \cite{DF} estimated the {\it average} correlations by only taking
into account the very rare events, which, in our notation (12)
for the generalized correlation length $\xi_q$ corresponds to the 
limit $q\to\infty$. Consequently his result $\nu=2$ should an upper
bound of our $\nu_q$. What is denoted by {\it typical} correlations
in \cite{DF} seems to us more related to our averaged correlation
functions.

Our finite-size analysis leads to finite values for
the dynamical exponent $z$, which might be due to the small system sizes we
were confined to. However, for $j_2\rightarrow0$ we find
$z\rightarrow\infty$ in agreement with a RNG-picture \cite{DF}.
Furthermore we find that the order-parameter probability
distribution scales non-trivially at the critical temperatures.
We made this observation also in connection with the cumulants of
the probability distribution of correlation functions and we found 
a hierarchy of critical exponents for the generalized correlation lengths.
Despite these facts the numerical value $\beta/\nu$ for the
finite-size scaling of the averaged spontaneous magnetization
concurs with the prediction made in \cite{DF}.

All these phenomena merit further investigation (a more detailed
discussion on these results will be given elsewhere\cite{bigone}), 
especially with respect to the Griffiths singularities occurring already
at temperatures above $T_c$ \cite{DF,shankar}. Finally we would like
to mention that there is also a large overlap
of the scenario we have encountered here with what might occur in
two- and three-dimensional Ising spin glass in a transverse field
\cite{qsg2d,qsg3d}.

\acknowledgments
AC is grateful to G.\ Parisi, E.\ Marinari and A.\ Vulpiani for an
interesting discussion and
thanks the Institut f\"ur Theoretische Physik of the University of 
K\"oln for hospitality where part of this work was done.
HR would like to thank L.\ Mikheev and A.\ P.\ Young
for many extremely valuable discussions.
He acknowledges the allocation of about $10^5$ processor hours
on the transputer cluster Parsytec--GCel1024 from the Center 
of Parallel Computing (ZPR) in K\"oln. This work was performed 
within the SFB 341 K\"oln--Aachen--J\"ulich.

\begin{figure}
\caption{Scaling plot of the magnetization $M(L,L_\tau)$ 
for $j_2=0.1$ at $T_c$. It yields $z=1.65\pm0.05$ and 
$\beta/\nu=0.17\pm0.01$ The insert shows 
the scaling plot of the susceptibility $\chi(L,L_\tau)$ with
$\gamma'/\nu=2.3\pm0.1$ and $z$ as for $M$.}
\end{figure}

\begin{figure}
\caption{Scaling plot of the averaged cumulant $g_{\rm av}(L,L_\tau)$
for $j_2=0.1$ at $T=T_c$ with $z=1.55\pm0.05$.
}
\end{figure}

\begin{figure}
\caption{{\bf Insert:} 
Correlation function in the (imaginary) time direction $C(0,t)$
for $j_2=0.05$ at $T=T_c$. The system sizes are $4\times16$, $6\times32$,
$8\times64$, $12\times100$ and $16\times160$, i.e.\ their aspect ratio
is close to the maximum of the cumulant $g_{\rm av}$ and therefore
roughly constant. The full curve is a least square fit to $C(0,t)\propto
t^{-\eta_\perp}+(160-t)^{-\eta_\perp}]$ and yields $\eta_\perp=0.23\pm0.01$.
{\bf Left part:}
Correlation function in the space direction $C(r,0)$
for $j_2=0.05$ at $T=T_c$. The system size is $16\times160$, close
to the maximum of $g_{\rm av}$. 
The full line is a least square fit 
to $C(r,0)\propto\exp(-a\,r^{1/2})+\exp(-a\,(L-r)^{1/2})$ with $a=0.62$.
}
\end{figure}

\begin{figure}
\caption{
Scaling plot for the typical correlation length in the spatial direction 
$\xi_0^{-1}L_\tau^{1/z_0}$ versus $(T-T_c)L_\tau^{1/z_0\nu_0}$. The upper
data-set is for $j_2=0.05$, where $z_0=2.12$ and $\nu_0=0.7$ has been used,
the lower data-set is for $j_2=0.1$, where $z_0=1.74$ and $\nu_0=0.7$.
}
\end{figure}


\begin{references}

\bibitem{XXZ}
C.\ A.\ Doty and D.\ S.\ Fisher,
Phys.\ Rev.\ B {\bf 45}, 2167 (1992);
S.\ Haas, J.\ Riera and E.\ Dagotto,
Phys.\ Rev.\ B {\bf 48}, 13174 (1994);
K.\ J.\ Runge and G.\ T.\ Zimanyi,
cond-mat/9312002.

\bibitem{exp} 
W. Wu, B. Ellmann, T. F. Rosenbaum, G. Aeppli and D. H. Reich,
Phys. Rev. Lett. {\bf 67}, 2076 (1991); 
W. Wu, D. Bitko, T. F. Rosenbaum and G. Aeppli;
Phys. Rev. Lett. {\bf 71}, 1919 (1993).

\bibitem{mft2}
J. Miller and D. Huse, Phys. Rev. Lett. {\bf 70}, 3147 (1993);
J. Ye, S. Sachdev and N. Read, Phys. Rev. Lett. {\bf 70}, 4011 (1993).

\bibitem{mft1}
For earlier work on the mean-field theory if quantum spin glasses see:
Y. Y. Goldschmidt and P. Y. Lai, Phys. Rev. Lett. {\bf 64}, 2567 (1990)
and references therein.

\bibitem{oppermann}
For the mean-field theory of intinerant spin glass models see
R.\ Oppermann, Nucl.\ Phys.\ {\bf B} 401, 507 (1993).

\bibitem{Santos}
B.\ Boechat, R.\ R.\ dos Santos and M.\ A.\ Continentino,
Phys.\ Rev.\ B {\bf 49}, 6404 (1994);
M.\ A.\ Continentino, B.\ Boechat and R.\ R.\ dos Santos, preprint (1994).

\bibitem{qsg2d}
H. Rieger and A. P. Young, submitted to Phys. Rev. Lett. (1994).

\bibitem{qsg3d}
M. Guo, R. N. Bhatt and D. A. Huse, submitted to Phys. Rev. Lett. (1994).

\bibitem{DF}
D. S. Fisher, Phys. Rev. Lett. {\bf 69}, 534 (1992).

\bibitem{lev} For recent work on layered Ising models see
L. V. Mikheev and M. E. Fisher, Phys.\ Rev.\ B {\bf 49},
378 (1994) and references therein.

\bibitem{mccoy}
B. M. McCoy and T. T. Wu, Phys. Rev. {\bf 176}, 631 (1968);
{\bf 188}, 982 (1969); {\bf 188}, 1014 (1969)

\bibitem{zittartz}
P. Hoever, W. F. Wolff and J. Zittartz, Z. Phys. B {\bf 41}, 43 (1981).

\bibitem{shankar}
R. Shankar and G. Murthy, Phys. Rev. B {\bf 36}, 536 (1987).

\bibitem{binder} 
K. Binder and J. S. Wang, J. Stat. Phys. {\bf 55}, 87 (1989).

\bibitem{suzuki}
M. Suzuki, Progr. Theor. Phys. {\bf 56}, 1454 (1976).

\bibitem{private}
We would like to thank Lev Mikheev for a very enlightening
discussion on this point. See also L.\ Mikheev, in preparation (1994).

\bibitem{kleban}
P.\ Kleban and G.\ Akinci, Phys.\ Rev.\ B {\bf 28}, 1466 (1983).

\bibitem{bhayou}
R. N. Bhatt and A. P. Young, Phys. Rev. B {\bf 37}, 5606 (1988).

\bibitem{CPSV}
         A. Crisanti, G. Paladin, M. Serva and A. Vulpiani,
         Phys. Rev. E, in press (1994).
         
\bibitem{Derrida}
         B. Derrida and H. J. Hilhorst
         J. Phys. C {\bf 14}, L544 (1981).

\bibitem{CNPV}
         A. Crisanti, S. Nicolis, G. Paladin and A. Vulpiani,
         J. Phys. A {\bf 23}, 3083 (1990).

\bibitem{crisanti}
	A. Crisanti, G. Paladin and A. Vulpiani,
	{\it Products of random matrices}, Springer series in 
	solid state sciences 104; Springer, Berlin--Heidelberg--New York 
	(1993).

\bibitem{bigone}
	A.\ Crisanti and H.\ Rieger, in preparation
         
         
\end{references}
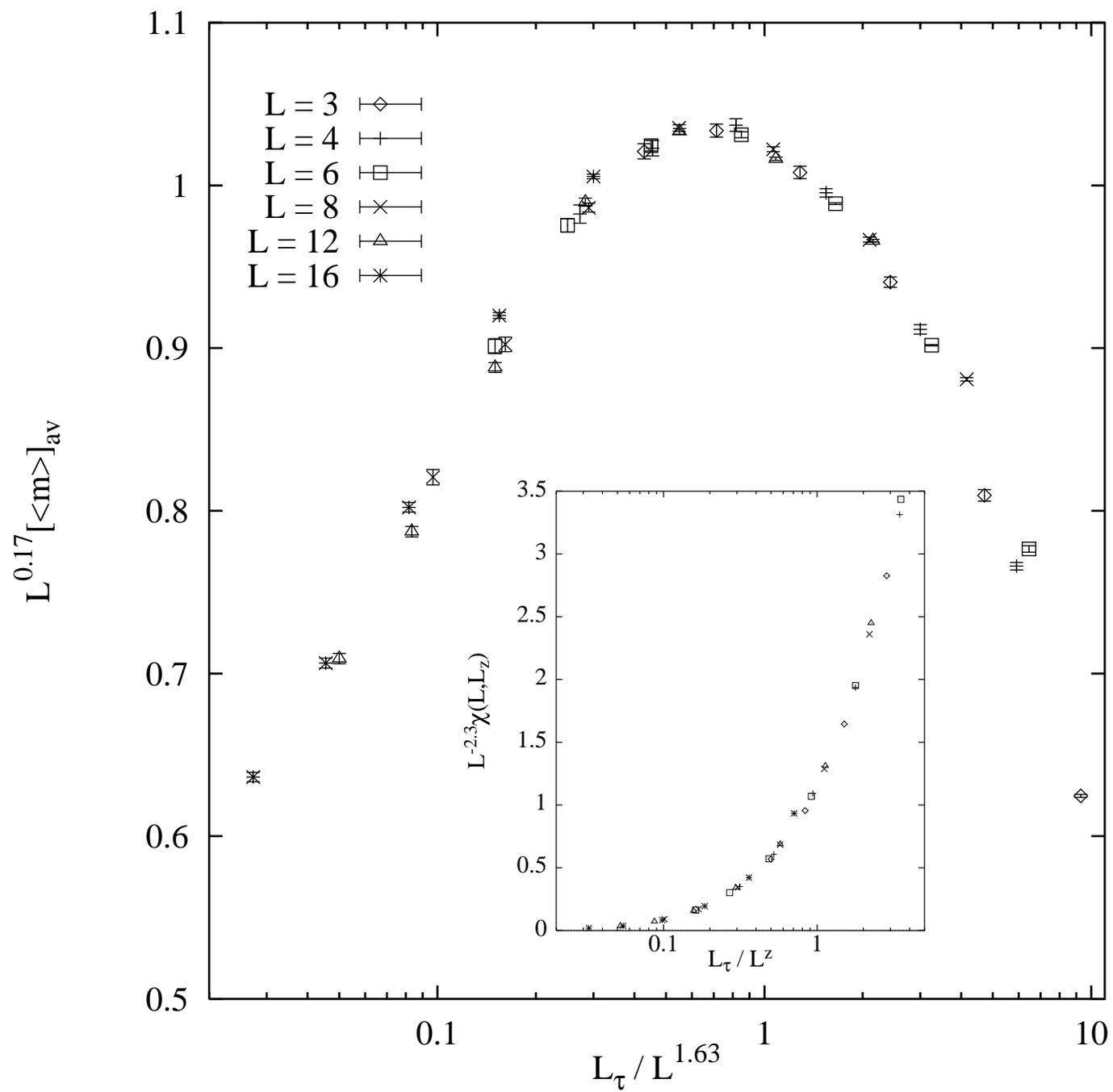
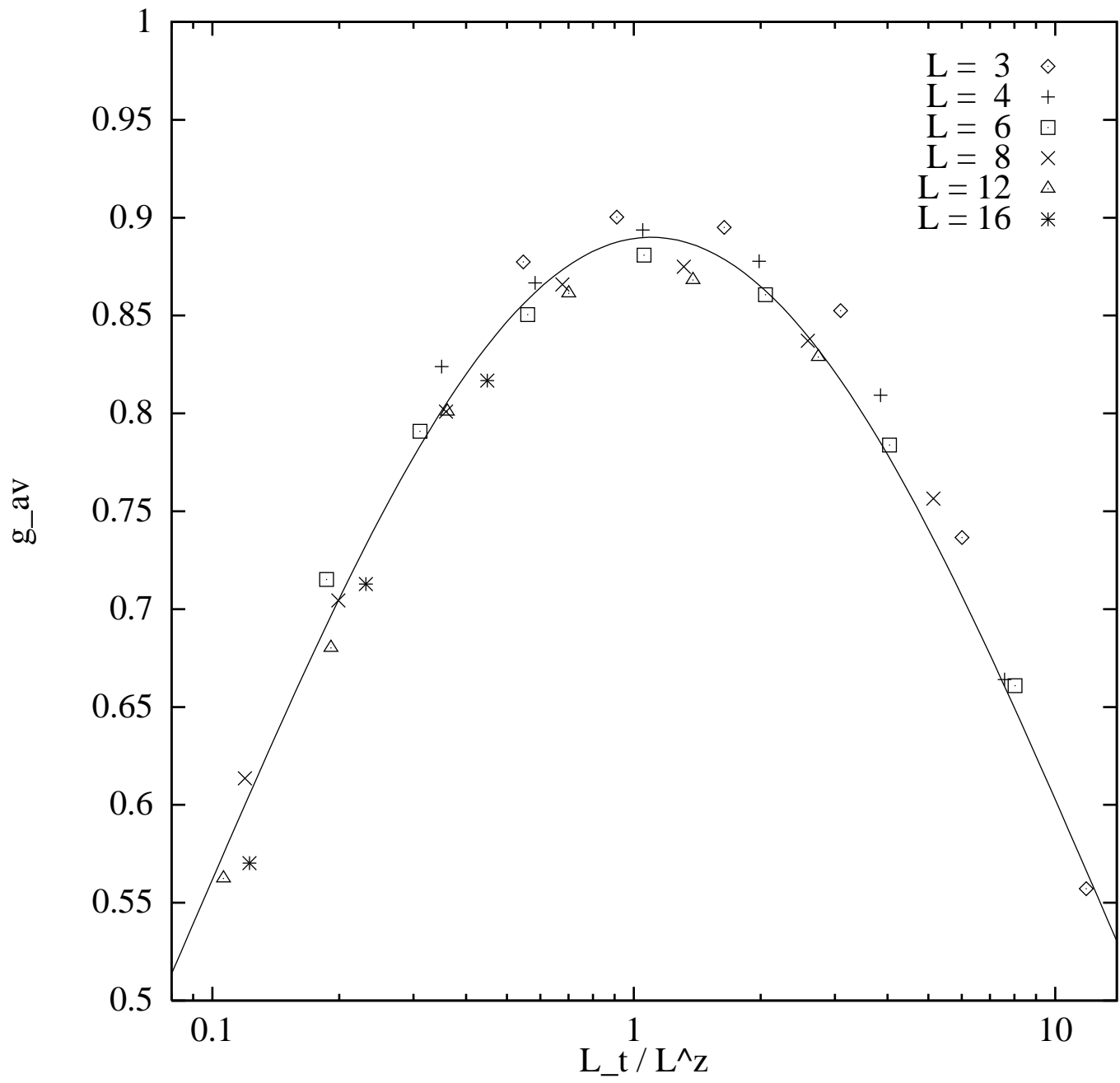
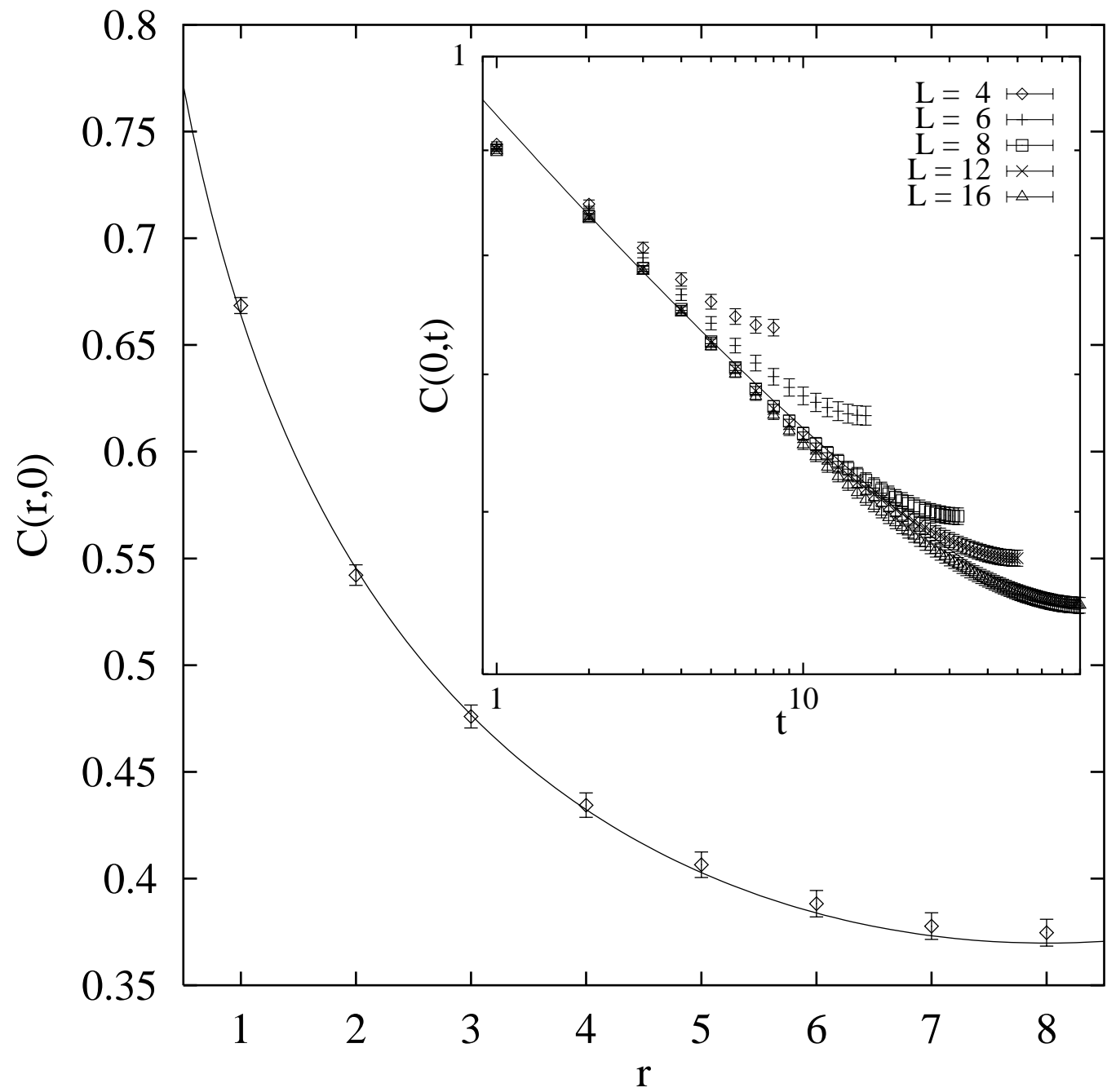
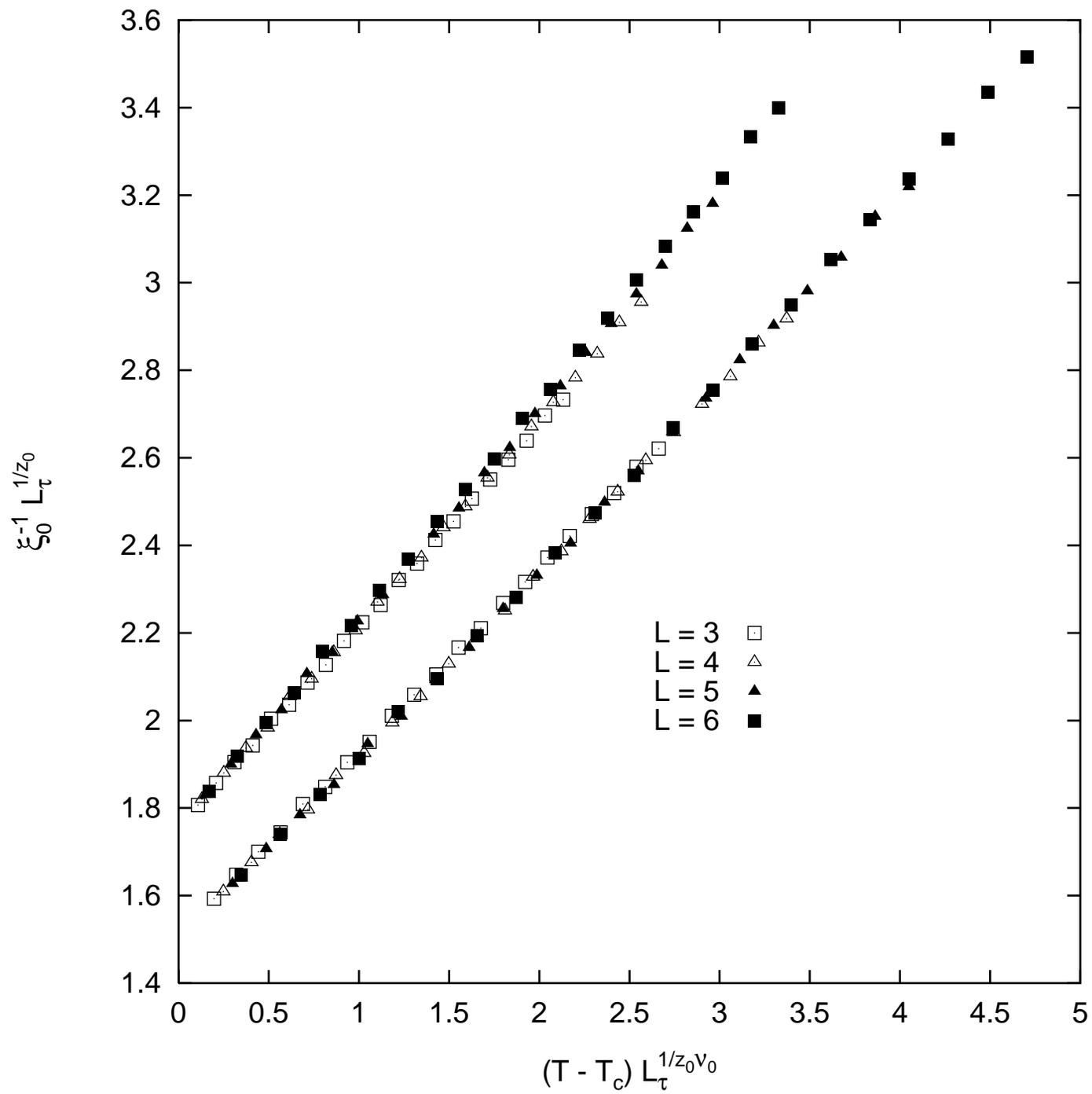
\end{document}